\documentclass{aa}
\usepackage{graphicx}
\usepackage[utf8]{inputenc} 
\usepackage{txfonts}
\usepackage{hyperref}       
\usepackage{url}            
\usepackage{booktabs}       
\usepackage{nicefrac}       
\usepackage{microtype}      
\usepackage{lipsum}
\usepackage{upgreek}
\usepackage{amssymb}
\usepackage[normalem]{ulem}
\usepackage{siunitx}
\usepackage[switch]{lineno}
\graphicspath{{./}{figures/}}
\usepackage{array}
\usepackage{ragged2e}
\newcolumntype{L}[1]{>{\raggedright\let\newline\\\arraybackslash\hspace{0pt}}m{#1}}
\title{First Results from the REAL-time Transient Acquisition backend (REALTA) at the Irish LOFAR station}
\titlerunning{REALTA First Results}
\authorrunning{Murphy et al.}

\author{
  P. C. Murphy
   \inst{1}\fnmsep\inst{2}
  \and
  P. Callanan
 \inst{3}
 \and
  J. McCauley
  \inst{1}
  \and
  D. J. McKenna
  \inst{2}\fnmsep\inst{1}
  \and
  D. \'O Fionnag\'ain
  \inst{4}
  \and
  C. K. Louis
  \inst{2}
  \and
  M. P. Redman
  \inst{5}
  \and
  L. A. Ca\~nizares
  \inst{2}\fnmsep\inst{1}
  \and
  E. P. Carley
  \inst{2}
  \and
  S. A. Maloney
  \inst{2}\fnmsep\inst{1}
  \and
  B. Coghlan
  \inst{6}
  \and
  M. Daly
  \inst{7}
  \and
  J. Scully
  \inst{7}
  \and
  J. Dooley
  \inst{3}
  \and
  V. Gajjar
  \inst{8}
  \and
  C. Giese
  \inst{1}\fnmsep\inst{8}
  \and
  A. Brennan
  \inst{5}\fnmsep\inst{8}
  \and
  E. F. Keane
  \inst{8}\fnmsep\inst{5}
  \and
  C. A. Maguire
  \inst{1}\fnmsep\inst{2}
  \and
  J. Quinn
  \inst{9}
  \and 
  S. Mooney
  \inst{9}
  \and
  A. M. Ryan
  \inst{1}\fnmsep\inst{2}
  \and
  J. Walsh
  \inst{6}
  \and
  C. M. Jackman
  \inst{2}
  \and
  A. Golden
  \inst{4}\fnmsep\inst{10}
  \and
  T. P. Ray
  \inst{2}
  \and
  J. G. Doyle
  \inst{10}
  \and
  J. Rigney
  \inst{2}\fnmsep\inst{10}\fnmsep\inst{11}
  \and
  M. Burton
  \inst{10}
  \and
  P. T. Gallagher
   \inst{2}\fnmsep\inst{1}
 }

\institute{School of Physics, Trinity College Dublin, Dublin 2, Ireland
\email{pearse.murphy@dias.ie}
\and
Astronomy \& Astrophysics Section, Dublin Institute for Advanced Studies, D02 XF86, Ireland.
\and
Department of Physics, University College Cork, Cork, Ireland
\and
Astrophysics Research Group, School of Mathematics, Statistics and Applied Mathematics, National University of Ireland Galway, University Road, Galway, H91 TK33, Ireland.
\and
Centre for Astronomy, School of Physics, National University of Ireland Galway, University Road, Galway, H91 TK33, Ireland.
\and
School of Computer Science and Statistics, Trinity College Dublin, Dublin 2, Ireland
\and
Deptartment of Computer and Software Engineering, Athlone Institute of Technology, Athlone, Ireland
\and
Department of Astronomy, University of California, Berkeley CA 94720, USA
\and
School of Physics, University College Dublin, Belfield, Dublin 4, Ireland 
\and
Armagh Observatory and Planetarium, College Hill, Armagh, BT61 9DG, N. Ireland
\and
Astrophysics Research Centre, School of Mathematics and Physics, Queen's University Belfast, Belfast BT7 1NN, N. Ireland 
}
\begin{document}

\abstract{Modern radio interferometers such as the LOw Frequency ARray (LOFAR) are capable of producing data at hundreds of gigabits to terabits per second. This high data rate makes the analysis of radio data cumbersome and computationally expensive. While high performance computing facilities exist for large national and international facilities, that may not be the case for instruments operated by a single institution or a small consortium. 
Data rates for next generation radio telescopes are set to eclipse those currently in operation, hence local processing of data will become all the more important.
Here, we introduce the REAL-time Transient Acquisition backend (REALTA), a computing backend at the Irish LOFAR station (I-LOFAR) which facilitates the recording of data in near real-time and post-processing. We also present first searches and scientific results of a number of radio phenomena observed by I-LOFAR and REALTA, including pulsars, fast radio bursts (FRBs), rotating radio transients (RRATs), the search for extraterrestrial intelligence (SETI), Jupiter, and the Sun.}

\keywords{instrumentation: miscellaneous, methods: data analysis, telescopes}

\maketitle

\section{Introduction}
\label{sec:intro}
Modern radio interferometers produce more data than any astronomical instrument in history. Interferometric observations can receive data at hundreds of gigabits per second, requiring high performance computer (HPC) facilities to preprocess data before it can be analysed and scientifically explored. With newly built telescopes such as the Murchison Widefield Array \citep[MWA;][]{Lonsdale2009}, MeerKAT \citep{Jonas2016} and the Australian Square Kilometre Array Pathfinder \citep[ASKAP;][]{Johnston2008} acquiring such vast amounts of data \citep[up to $300$~Gbps;][]{Voronkov2020, Lonsdale2009} radio astronomy is working at the cutting edges of big data science. The Square Kilometre Array \citep{McMullin2020} is commencing construction this year and pushes the computational requirements to ever more difficult regimes \citep{Scaife2020}.

The LOw Frequency ARray \citep[LOFAR;][]{VanHaarlem2013} is a radio interferometer located in eight countries across Europe. The majority of the LOFAR stations exist in a dense cluster in the Netherlands known as the core. There are a total of 52 LOFAR stations --- 24 core stations, 14 remote stations (also in the Netherlands), and 14 international stations (spread across Europe). The planned construction of an additional international station in Italy will increase this to a total of 53 LOFAR stations. The detailed differences between the station types are described in \cite{VanHaarlem2013}, one key difference being that international stations have twice the collecting area of a Dutch core or remote station. LOFAR operates in two modes: international mode, also known as International LOFAR Telescope (ILT) mode, and local, or stand-alone, mode. In ILT mode, data from all the LOFAR stations in the network are sent via fibre optics to the COBALT2.0 correlator \citep[an upgrade to COBALT1.0,][]{Broekema2018} in Groningen, in the Netherlands. In local mode, each international station is operated by the host institute or consortium. \textit{De novo}, international stations do not have dedicated processing backends. Without such an addition the raw data from an international station could not be recorded or analysed. 

\begin{figure*}[ht]
    \centering
    \includegraphics[width=0.8\linewidth]{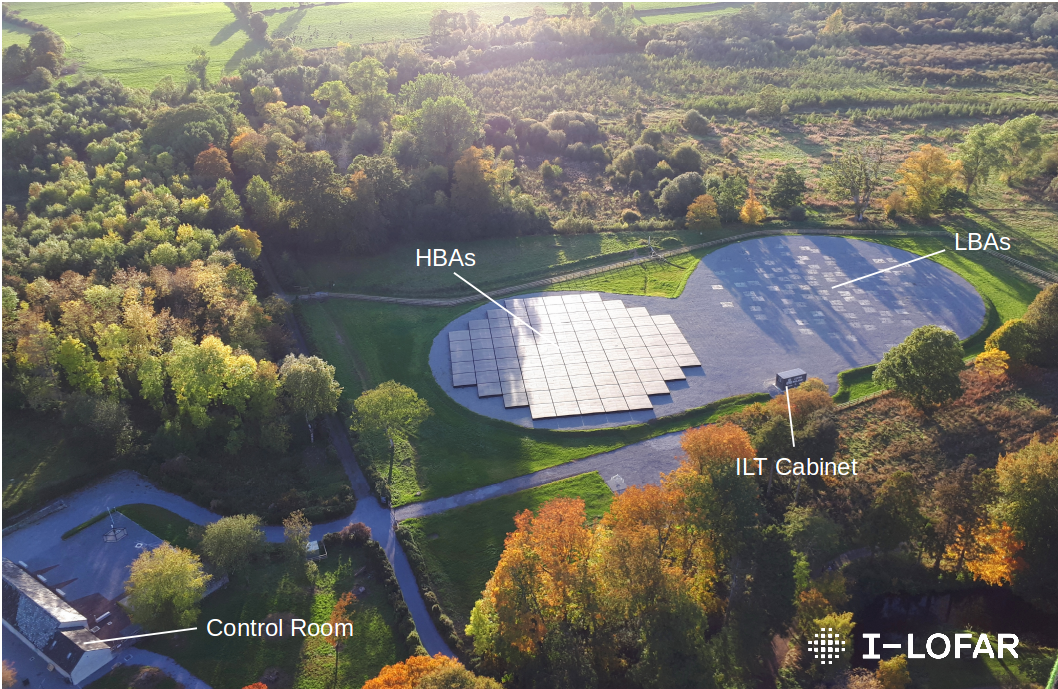}
    \caption{Aerial photograph of the Irish Low Frequency Array station IE613 (I-LOFAR) at Birr Castle, County Offaly. Data from the LBAs and HBAs are transferred to the ILT Cabinet (centre right) via coaxial cables where they are amplified, filtered and digitised. In international mode, data are transported to Groningen in the Netherlands at $\sim$3.2~Gbps. In local mode, data are processed using REALTA in the I-LOFAR Control Room (bottom left). Image credit: Alison Delaney (Birr Castle).}
    \label{fig:ILOFAR}
\end{figure*}

Construction of the Irish LOFAR station (I-LOFAR) was completed in July 2017 on the demesne of Birr Castle, County Offaly, in the Irish Midlands. An international LOFAR station such as I-LOFAR consists of $96$ dual-polarisation dipole antennas known as Low Band Antennas (LBAs) which record radio frequencies of \SIrange{10}{90}{\mega\hertz} and $96$ High Band Antenna (HBA) tiles which each contain $16$ bow-tie antennas connected to an analogue summator and record in the frequency range \SIrange{110}{250}{\mega\hertz} \citep[see][ for a full description of LOFAR antennas]{VanHaarlem2013}. Data recorded by the antennas are then channelised into $512$ subbands of \SI{195.3125}{\kilo\hertz} (\SI{156.25}{\kilo\hertz}) frequency resolution at \SI{5.12}{\micro\second} (\SI{6.4}{\micro\second}) temporal resolution depending on the clock rate used to sample data, either \SI{200}{\mega \hertz} or \SI{160}{\mega \hertz} \citep{VanHaarlem2013}. The signal from each antenna is then digitally beamformed (BF) to a direction on the sky to create beamlets. A beamlet is a specific location on the sky observed at a specific subband. An international LOFAR station can record and process data in either 8 bits or 16 bits, corresponding to a maximum of 488 or 244 beamlets respectively. The term to describe how many bits are used during an observation is the bitmode. It is also possible to record and process at 4 bits. Beamlets are recorded at $3.2$~Gbps. When I-LOFAR is in local mode, the BF data is sent along a fibre connection to a local control room. Fig. \ref{fig:ILOFAR} shows an aerial photograph of I-LOFAR. 

The total available bandwidth for I-LOFAR (and other international LOFAR stations) observations is determined by the bitmode used. 
This corresponds to a bandwidth of $\sim 47 \ (95, 190)$~MHz for 16 (8, 4) bit data. However, it is not necessary that the frequencies of each beamlet be sequential, which allows  a wider range of frequencies to be achieved. 
This is utilised in `mode 357', developed at the Kilpisjärvi Atmospheric Imaging Receiver Array \citep[KAIRA;][]{McKay-Bukowski2015}. In mode 357, beamlets are formed such that 200 beamlets from \SIrange{10}{90}{\mega\hertz}, 200 beamlets from \SIrange{110}{190}{\mega\hertz} and 88 beamlets from \SIrange{210}{240}{\mega\hertz} are recorded. In order to achieve this recording scheme, the number of antennas used for each frequency range is reduced. This leads to a lower sensitivity in mode 357. Mode 357 is particularly useful for observations of solar radio bursts, which typically occur across the entire LOFAR spectrum.

I-LOFAR produces a number of one second temporal resolution statistics files which are stored locally on the station's Local Control Unit (LCU). These are sub-band statistics (SSTs), which give the power spectrum for each antenna, beamlet statistics (BSTs), which give the power in each beamlet formed by the LOFAR station, and crosslet statistics (XSTs), the correlation coefficients between each antenna. These low-resolution data can be employed for system monitoring but are also sufficient for some astrophysical applications, for example, the XST data can be used to create snapshot all-sky images, and BST data have been used to study solar radio bursts \citep{Maguire2020}. The desire to use the full capabilities of the station, by accessing the full resolution is the motivation to develop a computational backend for the capture,  processing, and storage, of stand-alone BF data. 

The REAL-time Transient Acquisition backend (REALTA; from the Irish word for star, \textit{r\'ealta}) is a seven node computer cluster designed to record and analyse the raw BF data from international LOFAR stations in real-time, implemented at I-LOFAR. It takes inspiration from the ARTEMIS backend at the LOFAR-UK station in Chilbolton \citep{Serylak2012,Karastergiou2015}, although it significantly improves upon its hardware composition, using modern components with greater computational power. The REALTA hardware is available commercially, and as such REALTA can be implemented at any international LOFAR station and is ideal as a generic computing backend for LOFAR local mode observations.

Telescope backends at other international LOFAR stations have been used to search for fast radio bursts (FRBs) and studying pulsars using LOFAR-UK \citep[for example,][]{Karastergiou2015}, the French LOFAR station at Nançay \citep[for example,][]{Rajwade2016,Bondonneau2017}, and its extension, NenuFAR \citep[for example,][]{Bondonneau2020b}, the German stations \citep[e.g][]{Donner2019, Porayko2019, Tiburzi2019}, and the combination of a number of international stations \citep[for example,][]{Mereghetti2016, Hermsen2018, Michilli2018}. Here we showcase similar success in these and various other observations using I-LOFAR and REALTA. 


The hardware for REALTA is further described in \S~\ref{sec:REALTA} along with the networking configuration for the I-LOFAR control room while \S~\ref{sec:softwareAndProcessing} describes the data capture and processing software used by REALTA. In \S~\ref{sec:results}, some first scientific results from REALTA are highlighted. These include observations of pulsars, solar radio bursts, FRBs, Jovian radio emission, and the search for extraterrestrial intelligence (SETI).  

\section{REALTA}
\label{sec:REALTA}
\begin{figure*}[h]
    \centering
    \includegraphics[width=0.7\linewidth]{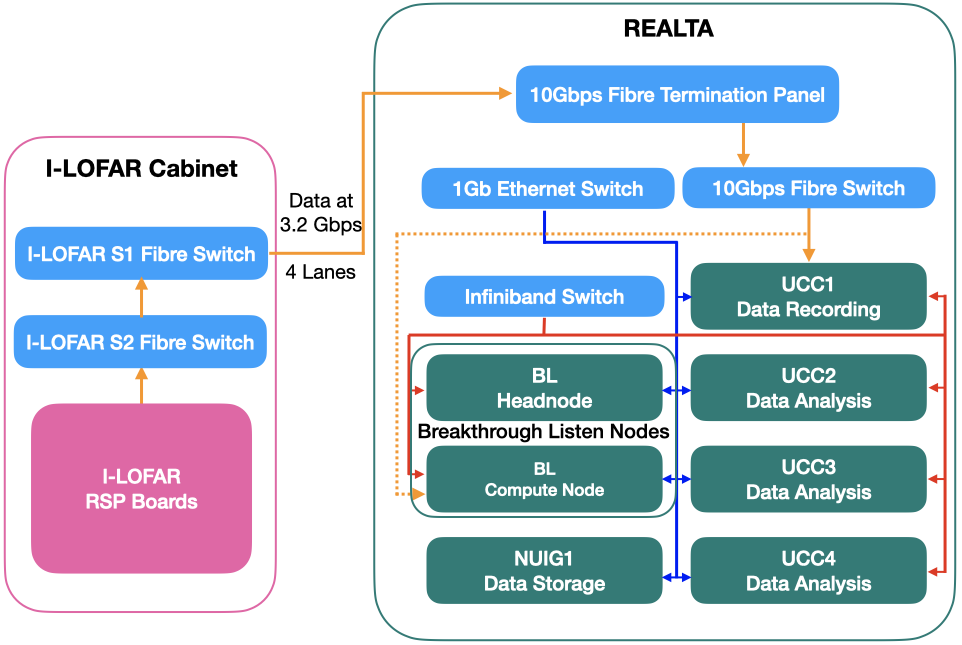}
    \caption{Block diagram for REALTA and I-LOFAR. Data recorded at the Remote Station Processing (RSP) boards are sent to the S1 fibre switch in the I-LOFAR container. Here the data are split into four `lanes' where each lane contains the data from a maximum of one quarter of the beamlets from the observation. The four lanes of data are then sent over a fibre connection to the I-LOFAR control room where it is recorded by REALTA. Orange arrows indicate the data path along fibre connections. Blue arrows are 1~Gbps Ethernet links and red arrows show infiniband connectivity. The dotted orange line is a fibre link to the BL compute node currently under development.}
    \label{fig:block}
\end{figure*}

In this section, we describe the local network configuration necessary to capture raw User Datagram Protocol \citep[UDP;][]{Postel} packets from I-LOFAR and the individual hardware components that make up REALTA. Data is transferred from the station through the UDP protocol using LOFAR CEntral Processing (CEP) packets. Each packet is 7824 bytes long and consits of a 16 bit header followed by 16 time slices for each beamlet. The full specification for CEP packets is described in both \cite{Lubberhuizen2009} and \citet{Virtanen2018}.  CEP packets are sent via fibre optics to CEP or REALTA over four data `lanes'. Each lane only holds one quarter of the maximum number of beamlets for an observation.
The network to facilitate capturing these packets is shown as a schematic block diagram in Fig. \ref{fig:block}, while Table \ref{table:REALTAspecs} gives a detailed description of the specifications for each of the REALTA nodes. 

\subsection{Local networking}
\label{sec:network}
The control room for I-LOFAR is located at the Rosse Observatory on the grounds of Birr Castle and $\sim$~100~m from I-LOFAR (Fig. \ref{fig:ILOFAR}). In order to record data from I-LOFAR while it is in local mode, a high-speed $10$~Gbps network was set up between the ILT container that houses the I-LOFAR Remote Station Processing (RSP) boards and the control room.

In a typical ILT observation, data output from each RSP board is sent to a Foundry LS648 10~Gbps network access switch (S2) in the container. From here it is sent to a Foundry LS624 switch (S1) before finally being sent over a 10~Gbps fibre connection to the COBALT cluster in Groningen for correlation and or beamforming with data from other LOFAR stations.

The aim of I-LOFAR's network configuration is to record RSP data to REALTA in the I-LOFAR control room. This is achieved by a fibre link between the I-LOFAR control room and S1. RSP data are sent along a Virtual Local Area Network (VLAN), on a 10~Gbps fibre link to the control room. Data reaches a fibre optic termination panel in the I-LOFAR control room and is sent to a 10~Gbps Dell EMC S4128F-ON optical switch. A fibre link between this switch and the REALTA compute nodes allows data to be recorded. All REALTA compute nodes are connected via Infiniband (an alternative to Ethernet and fibre). On REALTA this acts at a maximum of 10~Gbps to allow for transfer of data between nodes and Network File System (NFS) mounting.

As well as this, a future link to a HEAnet (Ireland’s national education and research network) cloud service for data transfer to the research institutions of the I-LOFAR consortium has been approved by ASTRON. This will allow access along existing 10~Gbps fibre infrastructure to S1 and then over another fibre to the control room along a VLAN. Another fibre is in place to eventually send RSP and Transient Buffer Board (TBB) data directly from S2 to the control room using two additional VLANs.
Data transfer from REALTA to research institutes is currently facilitated by a direct 1~Gbps between the Dublin Institute of Advanced Studies (DIAS) and S1 via HEAnet on a VLAN.

\subsection{Hardware description}
REALTA is ultimately designed to perform real-time analysis of radio data generated by the I-LOFAR international station. In its current form, it uses four Dell Poweredge R740XD compute nodes named UCC1, UCC2, UCC3, and UCC4. Each compute node contains two Intel Xeon\textsuperscript{\textregistered} Gold 6130 central processing units (CPUs) and an NVIDIA Tesla V100 16GB graphics processing unit (GPU). Each CPU has 16 cores, with two threads per core giving a total of 64 threads per node.  A total of $210$~TB of storage is distributed across the compute nodes with a further $128$~TB available on a dedicated storage server, NUIG1. Storage on REALTA is set up as a Redundant Array of Inexpensive Disks (RAID). Most disks ($\sim 263$~TB) are in RAID 5 for data archival, while a number of scratch disks ($\sim 75$~TB) are set up in RAID 0 for recording and processing raw data. In addition to this, a compute node for dedicated SETI research was provided by the Breakthrough Prize Foundation in collaboration with the Breakthrough Listen (BL) team at the Berkeley SETI Research Centre. The BL compute node is a SuperMicro 6049P-E1CR24H node with two, 16 core (8 threads) Intel Xeon\textsuperscript{\textregistered} Silver 4110 CPUs, an NVIDIA RTX 2080Ti GPU, and $144$~TB of storage. The BL Headnode is a SuperMicro 1029U-TRTP2 node intended to control REALTA during SETI observations. When being operated from the BL Headnode, all of the REALTA compute nodes will receive an identical operating system (OS) image and process in parallel across all five compute nodes. Each of the REALTA nodes connect to the 10~Gbps Dell EMC S4128F-ON optical switch via fibre optic cable as well as a 1~Gbps Ethernet switch for normal networking (using default VLAN 1). There are two redundant keyboard-video-mouse (KVM) servers and a KVM switch for remote access to both switches and all machines, and three uninterruptible power supplies (UPS) to mitigate the effects of a sudden power outage on both servers and critical access paths. Fail-safe cold-aisle air-conditioning will handle heat created by this unsupervised cluster.

REALTA is located in the control room which is $\sim 100$~m from the I-LOFAR HBAs and $\sim 150$~m from the LBAs. As REALTA was being set up and tested, the RFI from the system was monitored using all-sky observations and sub-band statistics observations. It was found that REALTA did not produce any significant RFI.

Table \ref{table:REALTAspecs} lists the technical specifications for each of the nodes in REALTA. In its configuration at the time of writing, REALTA acts as a number of independent nodes (UCC1-4, NUIG1, BL headnode, and BL Compute node). Data is recorded solely to the UCC1 compute node while the remaining nodes perform pre-processing and data analysis after data has finished recording. 

\begin{table*}\centering

\begin{tabular}{l L{4.5cm} L{4.5cm}}

  & Storage Node (NUIG1) & Compute Nodes ($\times 4$ UCC1-4 )  \\
 \hline
Machine Model & Dell Poweredge R730XD & Dell Poweredge R740XD \\
CPU  Model     & Intel Xeon\textsuperscript{\textregistered} E5-2640 V4 ($\times 2$)      &  Intel Xeon\textsuperscript{\textregistered} Gold 6130 ($\times 2$)       \\
CPU Clock Speed & 2.40GHz & 2.10GHz \\
CPU Cores (Threads) & 20 (40) & 32 (64)  \\
RAM       & 256GB        & 256GB          \\
Storage   & 128TB        & 210TB (total)          \\
GPU &  N/A & 16GB NVIDIA Tesla V100 \\
 & & \\
\end{tabular}

\begin{tabular}{l L{4.5cm} L{4.5cm}}

 & BL Headnode & BL Compute Node   \\
 \hline
Machine Model & SuperMicro 1029U-TRTP2 & SuperMicro 6049P-E1CR24H \\
CPU  Model    &  Intel Xeon\textsuperscript{\textregistered} Silver 4110 ($\times 2$)        & Intel Xeon\textsuperscript{\textregistered} Silver 4110 ($\times 2$)\\
CPU Clock Speed & 2.10GHz & 2.10GHz \\
CPU Cores (Threads) & 16 (32) & 16 (32) \\
RAM       & 93GB        & 96GB      \\
Storage   &  N/A        & 144TB      \\
GPU & N/A & 11GB NVIDIA RTX 2080Ti \\

\end{tabular}

\caption{Table of hardware specifications for REALTA. Note that the specifications are given for individual UCC 1-4 compute nodes, except for storage which is the total amount dedicated to archival of data distributed across all four. }
\label{table:REALTAspecs}
\end{table*}

\section{Software and pipelines}
\label{sec:softwareAndProcessing}
In REALTA's current state, UDP packets are recorded directly to disk before being converted into usable science products, such as Stokes parameters with variable levels of time integration, after the observation has completed. Future work will implement the current software pipelines to work directly on incoming data, which will allow for real-time analysis of the low frequency sky. Currently, a combination of local and community developed software is used in the main data recording and reduction pipeline. The primary recording and processing of the pipeline are described below.

\subsection{Recording data}
\label{sec:preproc}
The station RSPs generate a stream of packets across four data lanes, which is recorded on our compute nodes using software developed by Olaf Wucknitz at the Max Planck Institute for Radio Astronomy. The recording software, \texttt{dump\_udp\_ow}, listens to each port for incoming CEP packets, discards the UDP metadata related to the protocol, and writes the remainder of the packets to disk.

In order to reduce the storage requirements to record the CEP packets, the software can compress the data while it is being captured. This is accomplished by applying the \texttt{zstandard} compression algorithm\footnote{\hyperref[zstandard]{https://github.com/facebook/zstd}} to the output data stream. This procedure requires additional CPU processing but offers a compression fraction in the range of 40-60\%, depending on the observing mode and noise of an observation.

\subsection{Pre-processing data}
\label{sec:baseproc}
Once the data has been recorded to disk, observers use \texttt{udpPacketManager}\footnote{\hyperref[udpPacketManager]{https://github.com/David-McKenna/udpPacketManager}} \citep{McKenna2020}, a C library developed to convert the raw CEP packets, either compressed or uncompressed, to usable scientific data products. It implements checks to correct for issues that may have occurred during the recording process, such as padding when packets are missed, performs a polarmetric correction on the voltages using \texttt{dreamBeam}\footnote{\hyperref[dreamBeam]{https://github.com/2baOrNot2ba/dreamBeam/}}, and generates scientific data products for further processing and analysis.

The \texttt{udpPacketManager} library is extremely versatile and allows for an observer to chose a number of processing strategies. These include individual polarisation analysis, forming Stokes parameters, ordering data by either time or frequency or formatting data to be used with other software for further analysis. The most common processing strategy is to construct Stokes I and V from the voltages, then output the results to disk in a binary file format that follows the \texttt{SIGPROC} Filterbank standard \citep{Lorimer2011}. Future work will include determining best practises for removal of RFI from filterbank formatted files.


\subsection{Pulsar and single pulse processing}
\label{sec:pulsarproc}


All pulse-like observations undergo an additional processing step, whereby data are both channelised by a factor of eight (reducing the bandwidth of a given observing subband to 24.41~kHz) and coherent dedispersion is applied to the voltages.
Coherent dedispersion is the process by which an input signal is convolved with the inverse transfer function of the inter-stellar medium in order to remove the signature of dispersion delay due to free elections along the line of sight \citep{Hankins1987}. This process is especially important at frequencies observed by LOFAR as these delays scale with the inverse square of the observing frequency. This process is performed on the GPUs in REALTA using a modified version of \texttt{CDMT} \citep{Bassa2017}, which accepts input voltages from \texttt{udpPacketManager} rather than h5 files generated from the LOFAR COBALT system. In order to avoid distortions caused by Fast Fourier Transform (FFT) operations on zero-padded data, \texttt{CDMT} was further modified to overlap the input voltages between processing iterations.

Pulsar data is further reduced using a combination of \texttt{digifil} and \texttt{DSPSR} \citep{vanStraten2011} to generate folded profiles which are then analysed using \texttt{PSRCHIVE} \citep{Hotan2004}. The filterbanks are also often folded using \texttt{PRESTO} \citep{Ransom2001} to determine optimal dispersion measures for folding observations. RFI flagging is performed in two steps. Firstly, all data below 106~MHz and above 194~MHz is automatically flagged to remove contributions from local FM radio transmission and to minimise noise contributions due to the loss in sensitivity near the edge of the telescope's polyphase filter. Secondly, the spectral kurtosis method is performed by \texttt{DSPSR} during the folding step to remove transient RFI sources.

Single-pulse sources, such as intermittent rotating radio transients \citep[RRATs;][]{McLaughlin2006} and fast radio bursts \citep[FRBs;][]{Lorimer2007,Thornton2013} are searched for using \texttt{Heimdall}\footnote{\hyperref[Heimdall]{https://sourceforge.net/projects/heimdall-astro/}} to generate pulse candidates across a wide dispersion measure range. A typical search is performed between 5~pc~cm$^{-3}$ and 500~pc~cm$^{-3}$ across all pulsar and single-source observations. These are then filtered, discarding any below the $7.5~\sigma$ level to reduce the number of spurious candidates due to system noise or alignment of RFI between frequency channels. The remaining candidates are then plotted and visually inspected to discard those that are due to RFI, ionospheric scintillation or other phenomena that may cause spurious signals. 



\subsection{SETI data processing}
\label{sect:pipeline_SETI}
For SETI, the goal is to achieve a very high spectral resolution of the order of a few hertz to look for narrow-band Doppler drifting signals. Such signals are prime candidates for deliberately transmitted beacons by Extra-Terrestrial Intelligence \citep[ETI;][]{Tarter2001}. Baseband voltages are first recorded as described in \S~\ref{sec:preproc} directly on the BL compute node. The \texttt{udpPacketManager} library is then used to convert these data to Green Bank Ultimate Pulsar Processing Instrument\footnote{\hyperref[GUPPI]{https://safe.nrao.edu/wiki/bin/view/CICADA/GUPPiUsersGuide}} (GUPPI) formatted baseband data products for further processing. The BL team has developed a suite of software to work with the GUPPI formatted baseband voltages \citep{Lebofsky2019} and the preliminary result from this software is discussed in \S~\ref{sec:results}. 

\subsection{Future development and real-time analysis}
\label{sec:future_software}
\begin{figure*}[t]
    \centering
    \includegraphics[width=0.75\linewidth]{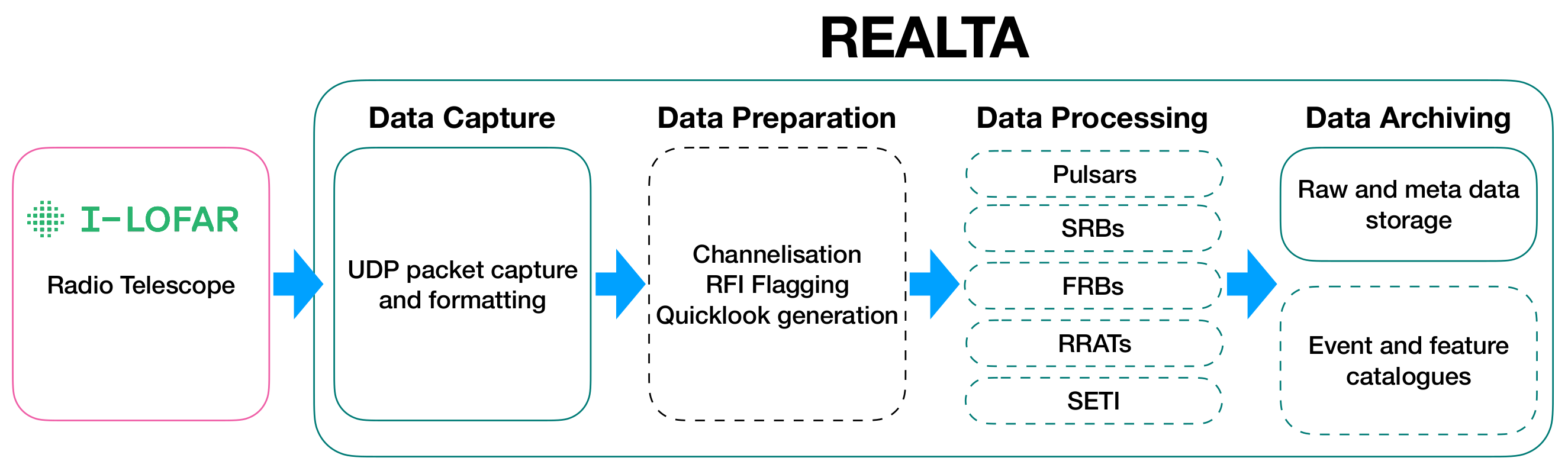}
    \caption{System diagram for REALTA, including future data preparation, processing, and archiving capabilities. Solid lines indicate existing features, while dashed lines denote stages under development. REALTA is currently capable of capturing data from the I-LOFAR radio telescope and archiving it in near real-time. We are in the process of developing a variety of data processing pipelines that will flag RFI, identify and characterise SRBs, pulsars, RRATs, FRBs, and SETI signals. Machine learning methods are being explored for a number of these tasks.}
    \label{fig:REALTA_future}
\end{figure*}

The current and planned data path through REALTA, from the time it is recorded at I-LOFAR to when it is written to disk by REALTA is shown as a block diagram in Fig. \ref{fig:REALTA_future}. UDP packets containing the data are captured and recorded directly to disk in real-time (\S \ref{sec:preproc}). The data are then formatted and metadata updated (\S \ref{sec:baseproc}) so that they are compatible with existing pulsar and SETI software (\S \ref{sec:pulsarproc}, \ref{sect:pipeline_SETI}).

In the future, the capture and formatting of data will occur simultaneously in real-time. Further channelisation of the raw data in the data capture stage will be implemented in order to increase the spectral resolution of observations and help in flagging Radio Frequency Interference (RFI) in the data. This, along with the generation of quick-look plots, will form the data preparation stage. The data processing software described below will also be developed to allow for real-time processing of solar, pulsar, FRB, RRAT, SETI, and other data. Finally, the data archive will be expanded to include a catalogue of transient events observed and a summary of their features.

\begin{figure*}[t]
    \centering
    \includegraphics[width=0.92\textwidth]{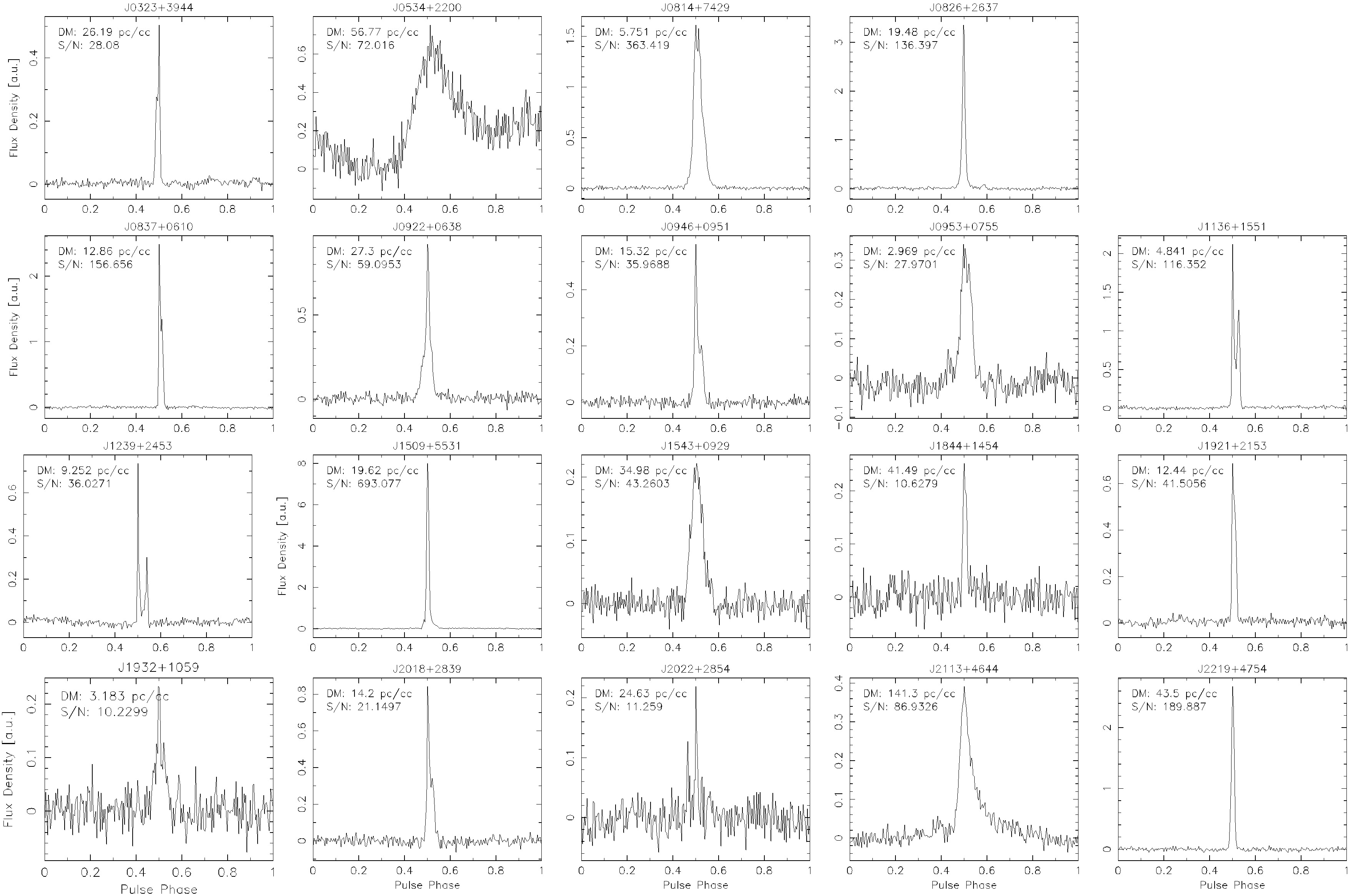}
    \caption{Sample of 19 pulsars observed with I-LOFAR and REALTA. Each of these observations is 6 minutes in duration and were taken on 4--5 March 2020 using the HBA antennas (\SIrange{110}{190}{\mega \hertz}). The data were processed using the method described in \S \ref{sec:pulsarproc} and plotted using \texttt{PSRCHIVE}. The x-axis in each plot is the pulse phase in radians, while the y-axis is flux density in arbitrary units.}
    \label{fig:pulsar-pc1}
\end{figure*}

\section{First results}
\label{sec:results}
Since REALTA's installation in July 2018, more than 130 unique targets have been observed. These science use cases range from the Sun, to planetary bodies, to pulsars. This allows for I-LOFAR to be used in the pursuit of a number of science goals. These include: analysing the wide gamut of low-frequency transient phase space \citep{KeanePhaseSpace2018}, characterising pulsars and the characteristics of binary systems \citep{Manchester2017}, and observing solar activity and space weather \citep{Maguire2020}. 
REALTA has also enabled collaborations with other LOFAR station operators, most notably an ongoing very long baseline interferometry (VLBI) campaign performed with stations located in Germany, France, and Sweden \citep[in a continuation of ][]{Wucknitz2019}.

The REALTA use cases are complementary to the Key Science Projects of the ILT \citep[KSP; see][]{VanHaarlem2013}. For example, a single international station is well suited to the study of bright transient sources such as pulsars, rotating radio transients, fast radio bursts, solar radio bursts, Jovian radio emission, and SETI signals, where flexible scheduling can be an advantage \citep[for example,][]{Maguire2020, Morosan2019}. The flexible scheduling of international stations in local mode allows for projects that require a large amount of observing time or regular observations of the same object. International station teams can also use the station to develop and test novel observing campaigns and hardware and software systems \citep[for example,][]{Scully2021}. First results from the aforementioned science use cases are described below.

\begin{figure*}[t]
    \centering
    \includegraphics[width=0.92\textwidth]{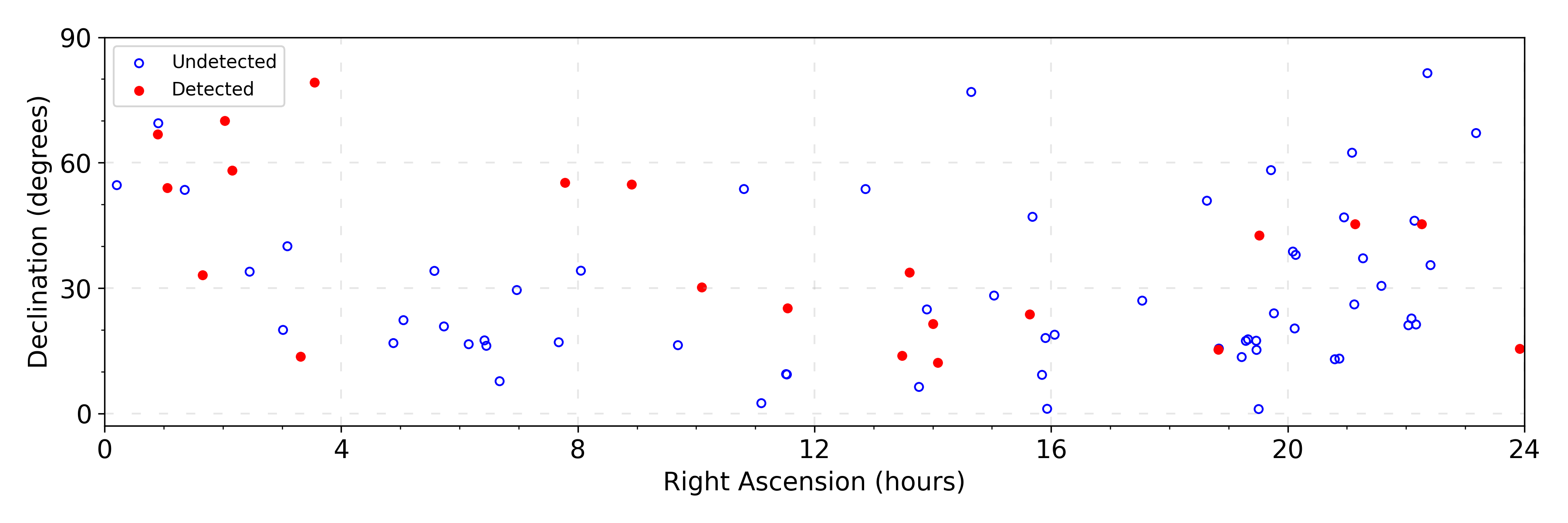}
    \caption{Overview of the sky positions of the RRATs observed during the census discussed in \S \ref{sec:rratsurvey} as of May 2021. Filled red dots indicate sources that have been observed and detected with either single pulses or periodic emission while blue circles indicate sources that were observed but not detected in I-LOFAR data.}
    \label{fig:rrat-pc1}
\end{figure*}

\subsection{Pulsars}

International LOFAR stations are ideal instruments to observe radio pulsars, particularly at frequencies between 100~MHz and 200~MHz \citep{Stappers2011, Bilous2014, Noutsos2015}. Further, the large fractional bandwidth of an international LOFAR station can offer new insight to the spectral variability of giant pulses from the Crab Nebula. Many international stations regularly participate in pulsar studies \citep[for example,][]{Mereghetti2016, Bondonneau2017, Hermsen2018, Donner2019}, while pulsar observations with LOFAR core stations \citep[for example,][]{Bilous2014, Bilous2020} are also common. Recent observations with the Polish international stations include those by \cite{Blaszkiewicz2020}. 

To date, over 50 different pulsars have been observed with I-LOFAR using REALTA. Fig. \ref{fig:pulsar-pc1} shows a sample of 19 pulsars which were observed, each for 6 minutes, on the 4th and 5th of March 2020, processed using the methods discussed in \S \ref{sec:pulsarproc}. In addition, as a result of regular timing campaigns of these sources, a number of targets have been studied in more depth, the Crab Nebula being the prime example. 

While the recording and timing of folded profiles of the Crab pulsar are of interest for studying the interior structure of the neutron star via its glitches \citep{Lyne2015} and variability due to scattering and echo events, it also frequently emits so-called `giant pulses' \citep{Meyers2017}. These giant pulses have fluences that vary from hundreds of Jy\,ms to tens of thousands of Jy\,ms. These pulses can be studied to analyse their scintillation, scattering, and brightness distributions.

\begin{figure}
    \centering
    \includegraphics[width=\columnwidth]{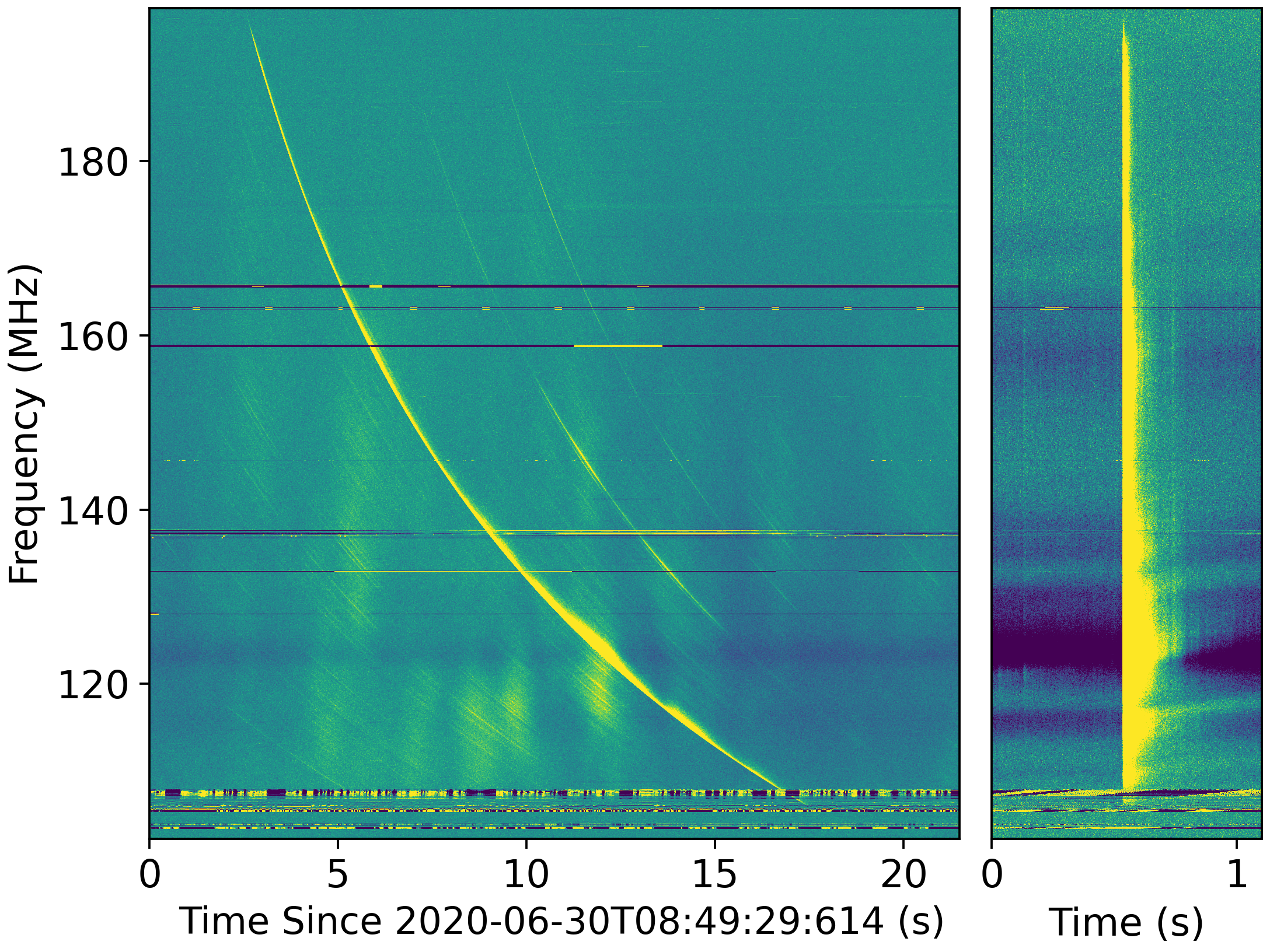}
    \caption{Observation of the Crab Pulsar performed on 30 June 2020, (left) without incoherent dedispersion and (right) with incoherent dedispersion. The plot on the left contains several giant pulses and both temporal- and spectral-variable ionospheric scintillation.  The plot on the right focuses on the brightest pulse in the group, which is the brightest pulse observed from the Crab Pulsar with I--LOFAR to date. Here data were processed to channel bandwidths of 24~kHz, resulting in a sampling rate of 40.96$\upmu$s. The data were time integrated to a temporal resolution of 1.31~ms prior to plotting.}
    \label{fig:giant_pulse}
\end{figure}

One such example of these giant pulses can be seen in an observation taken of the Crab pulsar with I-LOFAR on 30 June 2020. The observation was processed using the previously described methodology (\S \ref{sec:baseproc}, \S \ref{sec:pulsarproc}) on REALTA and a short segment of the observation is shown in Fig. \ref{fig:giant_pulse}. The figure is made from Stokes I data, where each channel has a bandwidth of 24~kHz and an underlying time resolution of 40.96~$\mu$s (though it has been interpolated using a median filter for this plot). There were $\sim 1300$ giant pulse candidates detected in the 30 minute observation. Using a rate of $\sim 0.7$ giant pulses per second, we determine there are $\sim 20$ giant pulses visible in the dynamic spectrum in the left panel of Fig. \ref{fig:giant_pulse}. Other structure in the left panel of Fig. \ref{fig:giant_pulse} includes some ionospheric scintillation and RFI. This time segment is mostly of interest due to one rare, extremely bright, giant pulse, which is re-plotted in the right panel of Fig. \ref{fig:giant_pulse} where it has been corrected for dispersion effects. Initial analysis using the radiometer equation indicates a specific fluence of $\sim 200$~kJy~ms across the observed bandwidth. Which may be the brightest pulse ever observed for this pulsar at these frequencies \citep{Karuppusamy2012,Meyers2017,VanLeeuwen2020}.

As of May 2021, over 80 hours of observations of the Crab pulsar have been captured by REALTA. Initial analysis of this data set, focusing on a four hour observation from March of 2020, has been performed and given results that are similar to that of other low-frequency instruments. The giant pulses were found to form a fluence distribution with a power-law fit of $\alpha = -2.86~\pm~0.07$, similar to that of other low frequency instruments \citep{Meyers2017}, but steeper than the results at Jodrell Bank \citep{Mickaliger2017}. Similarly, an initial investigation of the spectral behaviour of the scattering time scales of the brightest pulses appeared to follow a power law of $\alpha = -3.7 \pm 0.5$, in agreement with other instruments \citep{Driessen2019}. Future work includes planning to integrate a CLEAN-based de-convolution \citep[see][]{Bhat2003} of the pulse shapes to better describe the scattering and dispersion measure variations of single pulses over time, the effects of which are entangled \citep{McKee2018}, especially at lower frequencies.

\subsection{Rotating radio transients}\label{sec:rratsurvey}
Rotating radio transients (RRATs) are a class of neutron star that were discovered through detecting single, bright pulses rather than periodicity searches. If a sufficient number of pulses are detected within a short observing window, it is possible to determine the underlying period of the neutron star through brute force methods, after which the times of arrival of these pulses can be used to time the sources like any other pulsar \citep{Keane2011}. While the LOFAR core has blindly detected several RRATs during the LOTAAS survey \citep{Sanidas2019, Michilli2020, Tan2020}, and through targeted follow-up observations of sources detected with the Green Bank Telescope \citep{Karako-Argaman2015}, there has not been a major undertaking to time these sources with the LOFAR instruments, using the core or international stations.

However, the full-sky sensitivity and fractional bandwidth of a single international LOFAR station makes it the perfect candidate to perform follow-up observation on some of the brighter RRAT candidates identified by all-sky monitoring instruments such as the Big Scanning Array of Lebedev Physical Institute (BSA LPI) and the Canadian Hydrogen Intensity Mapping Experiment \citep[CHIME;][]{Amiri2018}. Follow-up observations of these candidates are useful to (a) determine the rotational characteristics of the stars through phase-coherent follow-up timing; and to (b) perform source characterisation from examining stars with broad spectral coverage.

Between July 2020 and May 2021, a 500 hour observing campaign has been undertaken to observe a diverse set of RRATs from the RRatalog\footnote{\hyperref[RRatalog]{http://astro.phys.wvu.edu/rratalog/}}, the CHIME-FRB Galactic sources database\footnote{\hyperref[CHIME-FRB Galactic Sources]{https://www.chime-frb.ca/galactic}}, and the BSA LPI Transients Catalogue\footnote{\hyperref[BSA LPI Transients Catalogue]{https://bsa-analytics.prao.ru/en/transients/rrat/}}, with a focus on sources that as yet do not have well defined periods. An overview of the sources observed and detected by this census can be seen in Fig. \ref{fig:rrat-pc1}. This campaign has so far resulted in the discovery of rotation periods for two sources which were previously unknown, periodic detection of a further two sources that have not been previously detected at LOFAR frequencies and the determination of coherent timing solutions for thirteen other sources. These results will be discussed in detail in a future paper (McKenna et al. in prep).

\subsection{Fast radio bursts}
Since their discovery in 2007 by \citeauthor{Lorimer2007}, fast radio bursts (FRBs) have been of keen interest to radio astronomers across the globe. While blind searches with multiple telescopes have helped push the lower bounds of their emission frequencies down year on year, the detection of numerous repeating FRBs by the CHIME-FRB collaboration \citep{CHIME2019} has accelerated this process in recent months. Searches for FRBs with LOFAR include those by \cite{Karastergiou2015} and \cite{TerVeen2019}, for example.

One particular repeating FRB, FRB 20180916B (`R3') has been found to have a period of $16.35\pm 0.15$ days, with an activity window of 5 days \citep{Amiri2020}, and has been detected with the LOFAR core as of December 2020 \citep{Pastor-Marazuela2020, Pleunis2021}. Prior to this, I-LOFAR and REALTA were used as a part of a 70 hour campaign to observe R3 during its activity phase and attempt to see emission at previously unseen frequencies. However, no significant pulse candidates were detected during this campaign. For further results and observations of other FRB sources see McKenna et al. (in prep).

\subsection{Solar radio bursts}
Solar radio bursts (SRBs) are some of the brightest phenomena in the radio sky. Five types of SRBs were classified in the 1950s \citep{Wild1950b, Boischot1957, Wild1959} and have been studied regularly since \citep[See][for a comprehensive review]{Pick2008}. A number of observations of solar radio bursts have been taken either using the LOFAR array as part of the Solar and Space Weather KSP \citep[for example,][]{Zhang2020, Murphy2021} or with an international station during local mode \citep[for example,][]{Morosan2019, Maguire2020, Bartosz2020}. Most solar radio bursts occur due to the plasma emission process, first described by \cite{Ginzburg1958}, and as such can be used as a diagnostic for the plasma density in the solar corona \citep{Melrose1987}. Remote sensing of radio emission from the Sun can be used as diagnostics of both large scale energy release from solar flares and CMEs \citep{Carley2021} and small-scale energy release, potentially related to coronal heating \citep{Mondal2020}.

\begin{figure}
    \centering
    \includegraphics[width=\columnwidth]{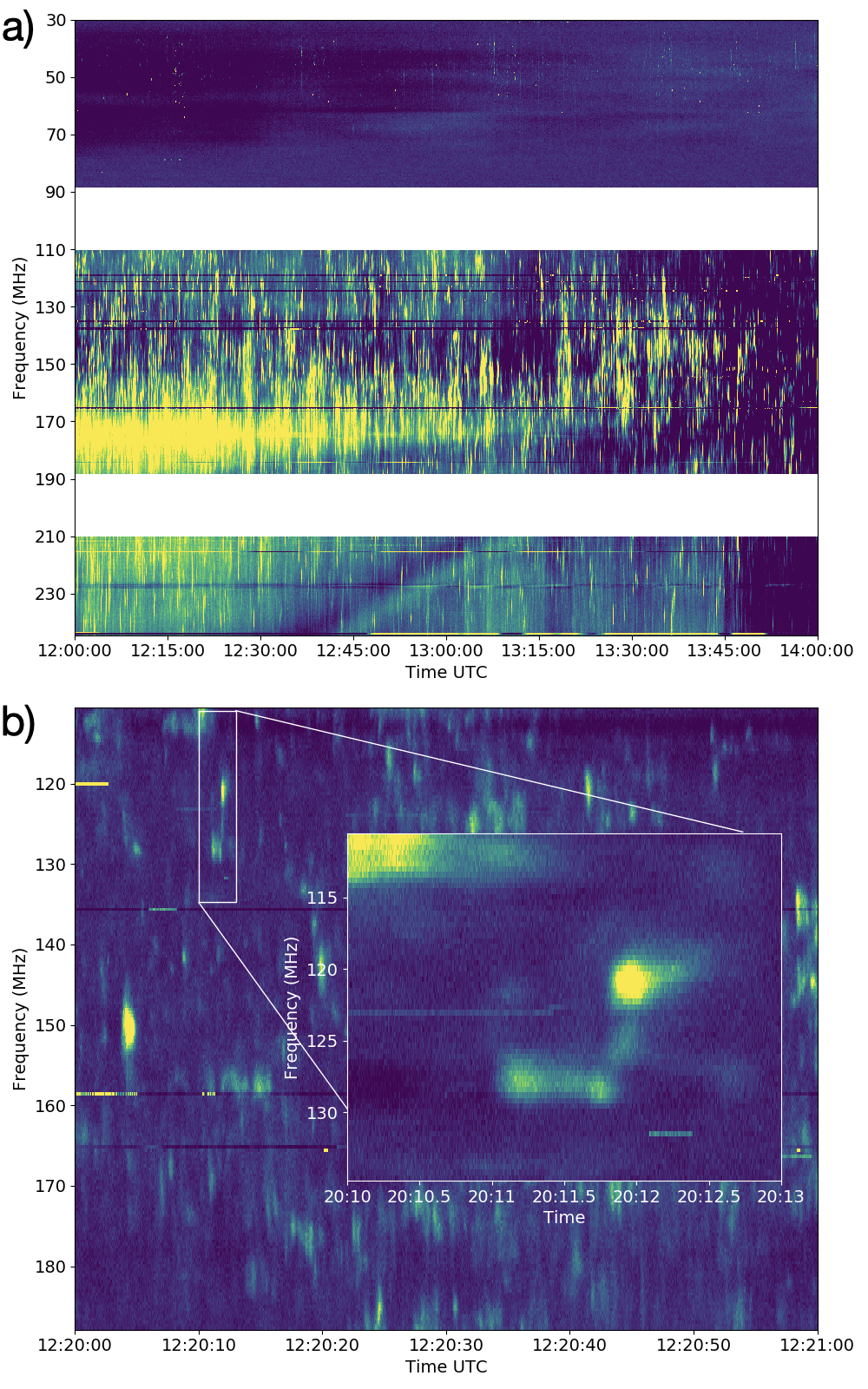}
    \caption{Solar radio noise storm observed on 2 November 2020. a) In this mode 357 observation, a number of bright bursts can be seen at frequencies greater than 110~MHz (yellow in the dynamic spectrum). 
    Here the data spans 2 hours from 12:00 UTC and has a temporal resolution of 10~ms.
    b) Zoom-in of panel a at 1~ms temporal resolution. A number of short duration SRBs are observed. The inset shows the sub-second variation of an individual burst in the noise storm, also at a temporal resolution of 1~ms.}
    \label{fig:357_10ms}
\end{figure}

On 2 November 2020, I-LOFAR observed a solar radio noise storm in mode 357. The dynamic spectrum of this storm from 12:00 - 14:00 UTC is shown in Fig. \ref{fig:357_10ms}a at 10~ms temporal resolution and \SI{195.3125}{\kilo \hertz} spectral resolution. A large number of short duration SRBs are seen across the full HBA band. Fine scale temporal and spectral structure are thought to be indicative of the turbulent nature of the solar corona which could further enhance the diagnostic capability of SRBs \citep{Kolotkov2018, Sharykin2018b, Reid2021}.

\begin{figure}
    \centering
    \includegraphics[width=\columnwidth]{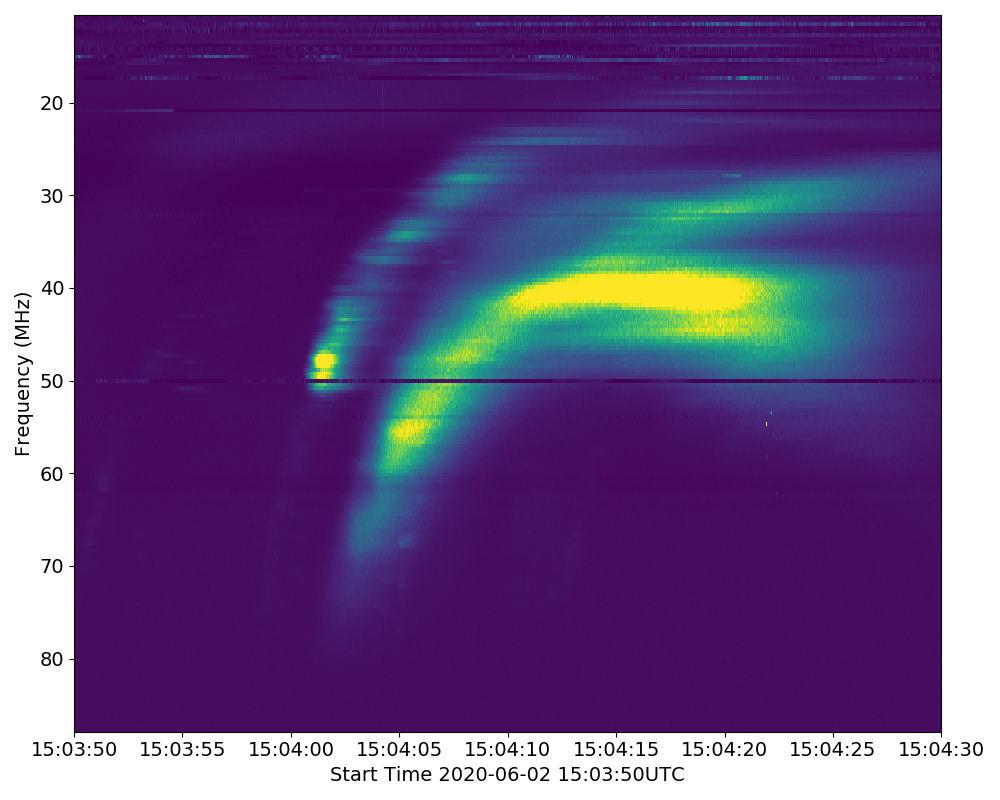}
    \caption{Two solar radio bursts observed on 2 June 2020. The earlier burst is likely a Type III SRB while the later burst shows the morphology typical of a U burst. The observation has a spectral resolution of $\sim$ \SI{195}{\kilo \hertz} and a temporal resolution of \SI{1}{\milli \second}.}
    \label{fig:uburst}
\end{figure}

Some SRBs can exhibit fine scale temporal and spectral features. These include, for example, herringbone bursts which are observed as part of Type II bursts \citep[for example,][]{Carley2015} or individual striations of a Type IIIb burst \citep[for example,][]{Zhang2020}. 
A number of short duration radio bursts, which are not part of the five classified types, have also been reported \citep[for example,][]{Ellis1967, Ellis1969, Melnik2010}. 
The high temporal resolution of REALTA observations with I-LOFAR will allow the study of these bursts at some of the highest temporal resolutions to date. Fig. \ref{fig:357_10ms}b shows a zoom-in of the radio noise storm from Fig. \ref{fig:357_10ms}a at 1~ms temporal resolution with an inset showing the sub-second variation of a particular burst. Fig. \ref{fig:uburst} shows an LBA observation from 2 June 2020 of a Type III burst and a U burst, both described by \cite{Reid2014}, for example, as being generated by electron beams travelling along open and closed magnetic field lines away from the sun respectively. This observation also has a 1~ms temporal resolution. Although a single international LOFAR station can only make spectroscopic observations of Type III bursts and their sub-types, this is still a valuable tool in determining the characteristics of the accelerated electron beam that instigates plasma emission in these bursts \citep{Reid2018}. Further, short duration pulsations in radio bursts can give insight into magnetohydrodynamic oscillations in the solar corona \citep{Carley2019}.

\subsection{Jovian auroral radio emission}

\begin{figure*}[t]
    \centering
        \includegraphics[width=0.8\textwidth]{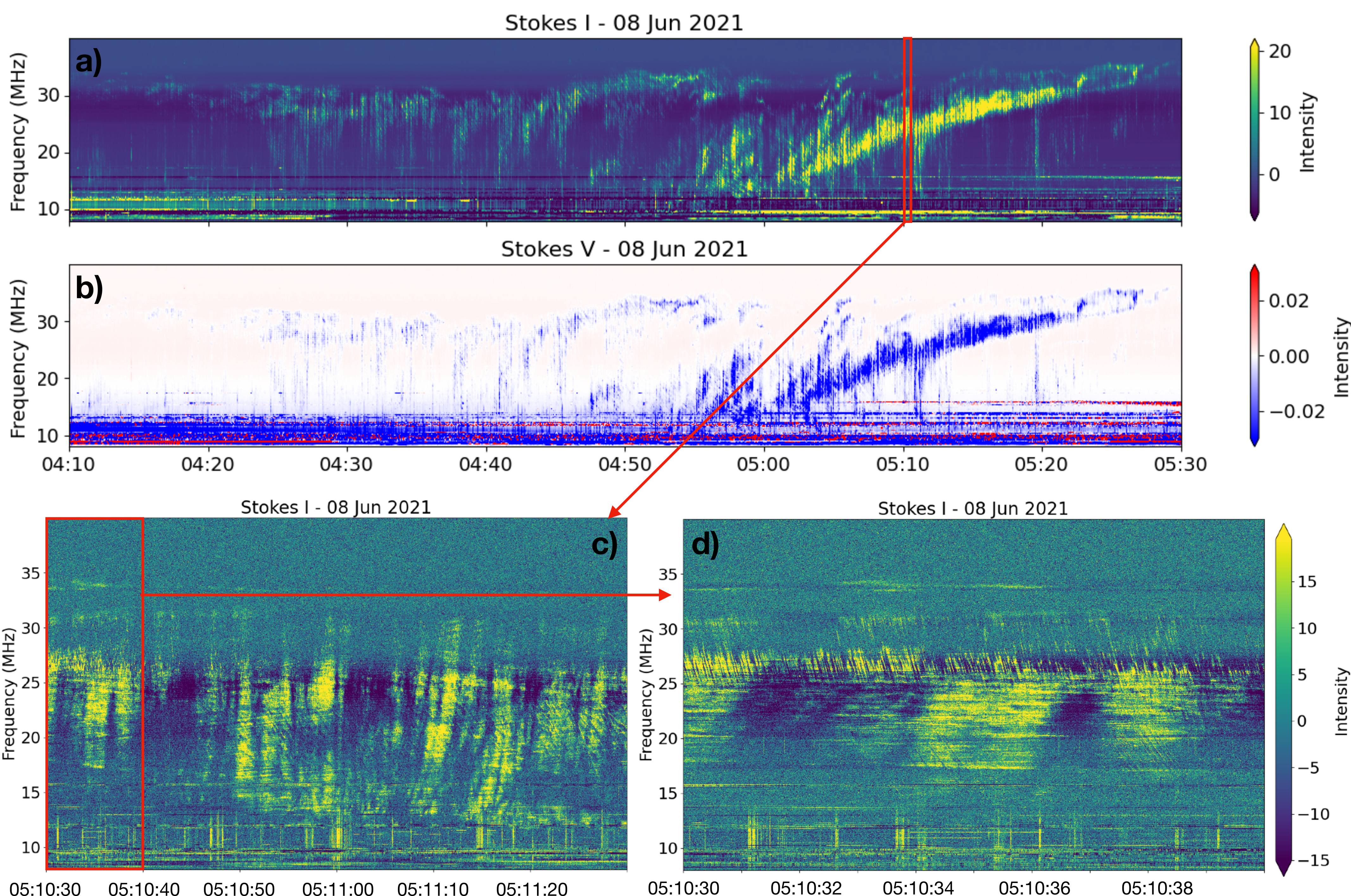}              
    \caption{Observation of a Jovian Decametric emission produced by the Io-Jupiter interaction. Panel (a) shows Stokes I (in dB above background). Panel (b) displays Stokes V (in arbitrary units). The resolution is $84$~ms per $12.2$~kHz. The emission displays a strong negative value in Stokes V, which means a strong right-hand circular polarization. Panels (c) and (d) show respectively a $60$~s and $10$~s zoom-in of panel (a) (Stokes I), processed with the highest resolution available for this observation ($81.92$~$\mu$s per $12.2$~kHz). Millisecond drifting bursts are visible panel (d).}
    \label{fig:io-stokesI}
\end{figure*}

Decametric radio emission (DAM), the strongest component of Jovian auroral radiation, was discovered in 1955 by \cite{Burke1955}, with part of this emission controlled by the Io-Jupiter interaction \citep{Bigg1964}. The source of this radio emission is known to be due to the electron cyclotron maser instability (ECMI) in the Jovian magnetosphere, which occurs when a circularly polarised wave resonates with the gyration of electrons with relativistic energies \citep{Wu1979,Wu1985,Zarka1998, Treumann2006,Louarn2017GRL}. The ECMI amplifies the wave on the extraordinary R-X mode which can escape the source and propagate in free space as a radio wave, at a frequency very close to the local electron cyclotron frequency, which is proportional to the local magnetic field amplitude. Jovian DAM emissions are the only planetary radio emissions visible from the ground, since part of the DAM is emitted at a frequency above the ionospheric cutoff frequency ($\gtrsim 10$~MHz). BF observations of Jovian DAM using the LOFAR core stations have been used to test the sensitivity of LOFAR to exoplanetary radio emissions \citep{Turner2019, Turner2021}.


On 8 June 2021, I-LOFAR observed Jovian DAM emission produced by the Io-Jupiter interaction, from 04:10 to 05:30 UTC. The Stokes I and V data from this observation are shown in Fig. \ref{fig:io-stokesI}a-b. An arc shape emission with a high intensity is observed between $\sim 04$:$55$ and $05$:$30$ (corresponding to the main Io-DAM emission), preceded by emissions with lower intensity starting at $\sim 04$:$10$ (corresponding to secondary Io-emissions). Looking at both the shape of the emission and its polarization (strong negative Stokes V value, corresponding to a right-handed circular polarization), we can determine that this emission is an Io-B emission \citep[coming from the north-dawn side of Jupiter, see][for example]{Marques2017}.

Moreover, we have with I-LOFAR access to very high temporal and frequency resolution, of which an example is shown Fig. \ref{fig:io-stokesI}c,d ($81.92$~$\mu$s per $12.2$~kHz). This will allow us to study the microphysics of the Jovian decametric emissions, for example the millisecond bursts visible Fig. \ref{fig:io-stokesI}d with a drifting feature in frequency with time ($\sim 25$-$30$~MHz). These millisecond drifts are thought to be electron bunches propagating along the magnetic field lines and can reveal both the energy of the resonant electrons as well as the potential drops (if present) along these fields lines \citep{Hess2007, Hess2009}. The high-resolution capability will also enable constraints to be placed on the position and movement of the sources, by interferometric measurements with several LOFAR stations, as well as the characteristics of the emission (for example, thickness and opening of the emission beam). Finally, I-LOFAR, combined with REALTA, is equipped to join the other LOFAR stations in the ground radio observation campaigns in support of the Jupiter space missions, both current (Juno\footnote{Such as the Juno Ground Radio Observation Support \hyperref[Juno]{https://maser.lesia.obspm.fr/task-1-data-collections/juno-ground-radio/}}) and future (JUICE, Europa-clipper).



\subsection{SETI}
\begin{figure}
    \centering
    \includegraphics[width=\columnwidth]{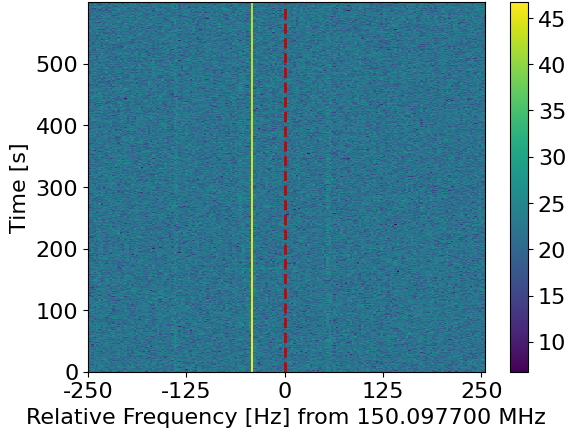}
    \caption{Narrowband signal detected using REALTA and the \texttt{turboSETI} algorithm. Dynamic spectra of an example narrowband signal detected in an observation pointing toward PSR B2217+47 using \texttt{turboSETI} with the BL nodes in REALTA. The colour bar shows intensity in arbitrary units while the red dotted line shows a relative frequency of 0~Hz from 150.0977~MHz. The signal does not show any drifting and thus likely has a terrestrial origin.}
    \label{fig:BL_fig}
\end{figure}

International LOFAR stations such as I-LOFAR have very broad fields of view, particularly at frequencies less than 150~MHz \citep{VanHaarlem2013}. This, coupled with the ability to channelise data to bandwidths $\lesssim 1$~Hz, are favourable characteristics in SETI research.
BL is conducting one of the most sensitive, comprehensive, and intensive searches for technosignatures on other worlds across a large fraction of the electromagnetic spectrum \citep{Worden2017}. Targets of the BL program include one million nearby stars, one hundred nearby galaxies, the entire Galactic plane, and exotic astrophysical objects \citep[see][for detail]{Isaacson2017}. \cite{Gajjar2019} provides the current status of these observing campaigns, as well as listing a number of collaborative observing facilities that are working alongside BL for carrying out these sensitive studies. The BL program is collaborating with two of the international LOFAR stations: I-LOFAR and LOFAR-SE, which is located at Onsala (Sweden), to complement searches towards the above-mentioned BL targets at lower radio frequencies. Details of the dedicated hardware deployed at I-LOFAR is discussed in \S \ref{sect:pipeline_SETI}. First-light observations were conducted with these BL nodes on 19 November 2020 towards PSR B1919+21 to validate the BL recording and conversion pipelines.
Recently, we also conducted observations of PSR B2217+47 on 21 April 2021 using the BL nodes for further pipeline development. Baseband data in the GUPPI format were converted to two different temporal and spectral resolution total intensity \texttt{SIGPROC} formatted filterbank data products. PSR B2217+47 was clearly detected, by folding high-temporal products. To search for narrowband Doppler drifting signals, \texttt{SIGPROC} formatted filterbank files with 3~Hz spectral resolution were used. This made use of the BL narrowband signal search tool, \texttt{turboSETI} \citep{Enriquez2017}. Fig. \ref{fig:BL_fig} shows an example of one of the narrowband signals of terrestrial origin detected using \texttt{turboSETI} towards PSR B2217+47. In the future, it is planned to conduct detailed on-target and off-target observations to discriminate such anthropogenic signals from true sky-bound ETI signals.

\section{Conclusions and future work}
\label{sec:future_work}

We have described the hardware for REALTA and given an overview of the software used to record and analyse data recorded from I-LOFAR. Several first result observations were showcased, exhibiting the broad range of objects that I-LOFAR and REALTA can observe. 

LOFAR 2.0 is a series of hardware and software upgrades to the ILT, which will be implemented in a number of stages over the coming years. An upgrade to an international LOFAR station, such as I-LOFAR, will require new receiver units (RCU2), new station beamformers (Station Digital Processors) based on the Uniboard$^2$ architecture \citep{Schoonderbeek2019} and new power, clock and control board for improved station control. The upgrade will greatly improve the instantaneous bandwidth, sensitivity and RFI rejection of an international station.


LOFAR for Space Weather (LOFAR4SW) \footnote{\hyperref[LOFAR4SW]{https://www.lofar4sw.eu}} is a proposed upgrade to LOFAR, currently being designed to enable regular space weather monitoring. If completed, LOFAR4SW would allow near-real-time monitoring of space weather phenomena such as solar flares and coronal mass ejections, interplanetary scintillation and ionospheric disturbances \citep{Carley2020a}. This is useful not only to space weather researchers but to the radio astronomy community as a whole as it will broaden our understanding of how space weather can effect the propagation of radio waves in the inner heliosphere and disturbances in the ionosphere and the effect this has on observing astronomical sources. In order to record the data streams from a LOFAR4SW enabled international station in local mode, a backend such as REALTA will be required to capture the data-stream and to process the raw data so that it can be used by space weather researchers and forecasters. The effectiveness of machine learning algorithms to detect solar radio bursts with REALTA is currently being investigated.

In the future, REALTA will be upgraded to fully include the BL headnode into the system. In order to achieve this, additional VLAN fibre connections will be set up between the UCC compute nodes and the BL headnode and compute nodes. Activating these VLAN fibre connections will allow each machine to record one lane of data and perform real-time channelisation and dedispersion making use of their GPUs.  The BL headnode will distribute an identical OS to the UCC and BL compute nodes and control them in parallel. This will allow REALTA to monitor for radio transients such as those important to SETI research. Upgrades to the data preparation and processing stages of the REALTA data flow (Fig. \ref{fig:REALTA_future}) will see REALTA operating fully in real-time (see \S \ref{sec:future_software}). 

With future upgrades to ILT hardware coming in the first half of the 2020s \citep[mainly, LOFAR2.0;][]{Edler2021}, international LOFAR stations will require a dedicated high-performance backend to record data rates of $\sim$6.4~Gbps, should they wish to use the full capacity of the instrument in local mode. While some international stations have existing backends, REALTA offers a powerful backend that is well suited to the data rates of LOFAR2.0. Backends like REALTA, due to the use of commercially available hardware, straight-forward network configuration and freely available software, will make it possible for international LOFAR stations to capture and process raw data, and to undertake a wider variety of astronomical observations. As mentioned in \S \ref{sec:future_software}, in the future it will be possible to process and record raw data in real-time. This will be done by recording data to a ring buffer implemented with the Parkes-Swinburne Recorder Distributed Acquisition and Data Analysis software \citep[PSRDADA;][]{PSRDADA} and then reading from the ring buffer into \texttt{udpPacketManager}. This will further improve the capabilities of international LOFAR stations during local mode. Finally, cooperation and coordinated observations between international LOFAR stations becomes easier if using the same software and hardware for data capture and post-processing and will be more beneficial than each international station operating individually.

\begin{acknowledgements}
We thank the referee for their thorough review and helpful suggestions.
P.C.M and D.McK are supported by Government of Ireland Studentships from the Irish Research Council (IRC). D.Ó.F. is supported by a Government of Ireland Postdoctoral Fellowship from the IRC (GOIPD/2020/145). I-LOFAR received funding from Science Foundation Ireland (SFI), the Department of Jobs Enterprise and Innovation (DJEI). TPR would like to acknowledge support from the European Research Council under Grant No. 743029, Ejection, Accretion Structures in YSOs (EASY). The I-LOFAR consortium consists of Trinity College Dublin, University College Dublin, Athlone Institute for Technology, Armagh Observatory and Planetarium (supported through funding from the Department for Communities of the N. Ireland Executive), Dublin City University, Dublin Institute for Advanced Studies, National University of Ireland Galway and University College Cork. REALTA is funded by SFI and Breakthrough Listen. Breakthrough Listen is managed by the Breakthrough Initiatives, sponsored by the Breakthrough Prize Foundation. 

\end{acknowledgements}

\bibliographystyle{aa}
\bibliography{references}

\end{document}